\documentstyle[12pt]{article}
\newcommand{\be}{\begin{eqnarray}}
\newcommand{\ee}{\end{eqnarray}}
\newcommand{\bdm}{\begin{displaymath}}
\newcommand{\edm}{\end{displaymath}}

\begin{document}
\begin{center}
{\LARGE {Aspects of thick braneworlds: 4D gravity localization,
smoothness, and mass gap}}
\end{center}
\vskip 1cm
\begin{center}
{\bf \large {Alfredo Herrera--Aguilar\footnote{E-mail:
herrera@ifm.umich.mx}}},
\end{center}
\begin{center}
{\bf \large {Dagoberto Malag\'on--Morej\'on\footnote{E-mail:
malagon@ifm.umich.mx}}},
\end{center}
\begin{center}
{\bf \large {Refugio Rigel Mora--Luna\footnote{E-mail:
topc@ifm.umich.mx}}}
\end{center}
\begin{center}
and
\end{center}
\begin{center}
{\bf \large {Ulises Nucamendi\footnote{E-mail:
ulises@ifm.umich.mx}}}
\end{center}
\vskip 0.1cm
Instituto de F\'{\i}sica y Matem\'{a}ticas, Universidad
Michoacana de San Nicol\'as de Hidalgo. \\
Edificio C--3, Ciudad Universitaria, C.P. 58040, Morelia,
Michoac\'{a}n, M\'{e}xico. \\

%%%%%%%%%%%%%%%%%%%%%%%%%%%%%%%%%%%%%%%%%%%%%%%%%%%%%%%
\begin{abstract}

We review some interrelated aspects of thick braneworlds constructed
within the framework of 5D gravity coupled to a scalar field
depending on the extra dimension. It turns out that when analyzing
localization of 4D gravity in this smooth version of the
Randall--Sundrum model, a kind of dichotomy emerges. In the first
case the geometry is completely smooth and the spectrum of the
transverse traceless modes of the metric fluctuations shows a single
massless bound state, corresponding to the 4D graviton, and a tower
of massive states described by a continuous spectrum of
Kaluza--Klein excitations starting from zero mass, indicating the
lack of a mass gap. In the second case, there are two bound states,
a massless 4D graviton and a massive excitation, separated by a mass
gap from a continuous spectrum of massive modes; nevertheless, the
presence of a mass gap in the graviton spectrum of the theory is
responsible for a naked singularity at the boundaries (or spatial
infinity) of the Riemannian manifold. However, the imposition of
unitary boundary conditions, which is equivalent to eliminating the
continuous spectrum of gravitational massive modes, renders these
singularities harmless from the physical point of view, providing
the viability of the model.

\end{abstract}

%\keywords{Localization of 4D gravity, mass gap, thick branes, naked
%singularities.}

\section{Introduction}

In recent years, several results have been obtained in the framework
of thick braneworld scenarios with a single extra dimension
\cite{dewolfe}--\cite{Yang}. These models represent more realistic
generalizations of several pioneer thin brane configurations
\cite{akama}--\cite{farak}, since if there are extra dimensions,
they must be accessible at least at some hypothetic energy scale.

One way to smooth out the singular brane configurations is to
replace the delta functions in the action of the system by a
self--interacting scalar field. Once we have a smooth brane
configuration, it is important from the phenomenological point of
view to look for configurations that allow for the existence of a
mass gap in the spectrum of gravitational Kaluza--Klein (KK)
excitations. This gap fixes the energy scale at which massive modes
can be excited and it is relevant for experimentally distinguishing
the imprints of the massless zero mode, identified with a stable 4D
graviton, from those coming from the massive modes, either discrete
or continuous. When there is no mass gap in the spectrum, there are
several modes with very small masses that cannot be experimentally
distinguished from the massless one.

The relevance of braneworld models is related to the fact that we
could live in a higher dimensional world without contradiction with
up to day experimentally tested 4D gravitational effects.

An interesting issue is related to the fact that naked singularities
arise when a mass gap is present in the 5D spectrum of
transverse traceless modes of metric fluctuations when
studying 4D gravity localization on thick brane configurations. In
Sec. II it is shown that if one considers localization of 4D gravity
with a generic warped ansatz for the metric, the presence of a mass
gap in the spectrum of the transverse traceless fluctuations
necessarily implies the existence of naked singularities located at
the boundaries (or spatial infinity, depending on the used
coordinate system) of the fifth dimension or, alternatively, the 5D
manifold can be completely smooth iff the massive spectrum of KK
excitations is gapless. This analysis is made in terms of a simple,
but generic relationship existing between the 5D curvature scalar,
the warp factor and the quantum mechanical potential which governs
the dynamics of the linearized transverse traceless modes of metric
fluctuations. We follow a traditional point of view which states
that spaces with naked singularities are physically acceptable only
if one imposes unitary boundary conditions that guarantee 4D energy
and momentum conservation
\cite{gremm},\cite{GMZ},\cite{cohenkaplan}. By analyzing the
spectrum of KK modes when imposing unitary boundary conditions, one
can see that the continuum part of the spectrum is projected out.
Thus, the theory of the discrete part of the spectrum remains
unitary because these modes die off rapidly enough as we approach
the singularity, a fact that provides a viable model.

\section{4D gravity localization: smoothness vs. mass gap}

Let us consider the following 5D action \cite{dewolfe,gremm}
\begin{equation}
\label{action} S_5=\int
d^5x\sqrt{|G|}\left[-\frac{1}{4}R_5+\frac{1}{2}(\nabla\phi)^2-V(\phi)\right],
\end{equation}
where $\phi$ is a bulk scalar field and $V(\phi)$ is a
self--interacting potential for the scalar field. We shall study a
solution which preserves 4D Poincar\'e invariance with the metric
\begin{equation}
\label{conflinee} {ds}_5^2=e^{2A(y)}\eta_{nm}dx^n dx^m-dy^2,
\end{equation}
where $e^{2A(y)}$ is the warp factor, $m,n=0,1,2,3$.

In the general case the analysis of classical gravitational
perturbations in braneworld models may be complicated because the
metric fluctuations are coupled to the fluctuations of the scalar
field. On the other hand, the invariance of the theory under
diffeomorphisms makes the very definition of perturbations gauge
dependent, in other words, by performing a small--amplitude
transformation of the space--time  coordinates (a gauge
transformation), we can easily introduce ``fictitious" perturbations
in the background  space--time. There are two approaches to deal
with these gauge ambiguities. The first one is to fix a gauge, i. e.
to pick a coordinate system which completely  eliminate the gauge
freedom; the second one consists of extracting some gauge--invariant
variables to obtain coordinate independent results \cite{massimo}.
In our work we consider the gauge--invariant method.

If 4D Poincar\'e invariance is unbroken, the different modes of the
geometry should be classified according to 4D Poincar\'e
transformations in scalar, vector and tensor sectors \cite{massimo},
where the vector modes are divergenceless and the tensor modes are
transverse and traceless.

The scalar and  vector fluctuations of the geometry are not
invariant under a general 5D infinitesimal coordinate transformation
(i. e. gauge--invariant), so that suitable gauge invariant variables
can be chosen in order to study the perturbation equations in a
gauge--invariant framework \cite{massimo}. On the other hand, the
tensor modes are automatically invariant under a general 5D
infinitesimal coordinate transformation. Futhermore, in linear
theory of perturbations, there is no coupling between the different
fluctuation modes, and hence they envolve independently. Following
\cite{dewolfe}, we shall just study the transverse traceless modes
of the background fluctuations $h_{mn}^T$ since they decouple from
the sector of the scalar field perturbations.

Let us study the metric fluctuations $h_{mn}^T$ of (\ref{conflinee})
given by
\begin{equation}
\label{mfluct}
ds_5^2=e^{2A(y)}\left[\eta_{mn}+h_{mn}^T(x,y)\right]dx^m dx^n-dy^2,
\end{equation}
where $ \eta^{mn} h_{mn}^T=\partial^{m}h^T_{mn}=0$.

In order to get a conformally flat metric we perform the coordinate
transformation \begin{equation} \label{coordtransf} dz=e^{-A}dy.
\end{equation}
Thus, the equation that governs the dynamics of the transverse
traceless modes of the metric fluctuations $h_{mn}^T$ becomes
\cite{dewolfe,gremm}
\begin{equation}
\label{eqttm}
\left(\partial_z^2+3A'\partial_z-\Box\right)h_{mn}^T(x,z)=0,
\end{equation}
where $A'=dA/dz$. This equation supports a massless and normalizable
4D graviton given by $h_{mn}^T=K_{mn}e^{ipx}$, where $K_{mn}$ are
constant parameters and $p^2=m^2=0$.

We adopt the following ansatz
$$h_{mn}^T=e^{ipx}e^{-3A/2}\Psi_{mn}(z)$$ in order to recast
(\ref{eqttm}) into a Schr\"{o}dinger's equation form (here we have
dropped the subscripts in $\Psi$ for simplicity):
\begin{equation}
\label{schrodinger} [\partial_z^2-V_{QM}(z)+m^2]\Psi(z)=0,
\end{equation}
where $m$ is the 4D mass of the KK excitation modes, and the analog
quantum mechanical potential has the form
\begin{equation}
V_{QM}(z)=\frac{3}{2}\left[\partial_z^2A+\frac{3}{2}(\partial_z
A)^2\right]. \label{VQM}
\end{equation} As it is shown in
\cite{csakietal}, from (\ref{schrodinger}) it follows that the zero
energy wave function is given by
\begin{equation}
\label{Psi0} \Psi_{0}(z)=e^{\frac{3}{2}A(z)}.
\end{equation}
The condition for localizing 4D gravity demands $\Psi_{0}(z)$ to be
normalizable; in other words, the zero energy wave function must
satisfy the following relationship
\begin{equation}
\int dz |\Psi_{0}|^{2}<\infty. \label{norma}
\end{equation}
Finally, it is worth noticing that the curvature scalar
corresponding to the ansatz (\ref{conflinee}) in the $z$ coordinate
adopts the form
\begin{equation}
\label{R}
R_5=8e^{-2A}\left[\partial_z^2A+\frac{3}{2}\left(\partial_z
A\right)^2\right].
\end{equation}
As it was pointed out in \cite{csakietal,bs} in order to have a
localized massless mode of the 5D graviton, $V_{QM}$ must be a well
potential with a negative minimum which approaches a positive value
$V_{QM_\infty} \ge 0$ as $|z|\rightarrow{\infty}.$ This property
ensures the fulfilment of the requirement (\ref{norma}).

Moreover, if we wish to have a well defined effective field theory,
it is desirable to obtain solutions where the massless graviton is
separated from the massive modes by a mass gap \cite{bs}.
This phenomenological aspect ensures the lack of arbitrary light KK
excitations in the mass spectrum. If the quantum mechanical
potential asymptotically approaches a positive value
$V_{QM_\infty}>0$, the existence of a mass gap is guaranteed.

Let us determine how these two aspects are related to the smooth
character of the curvature scalar. Thus, we must study the relation
among the following aspects of the theory:
\begin{description}
  \item[a)] Localization of 4D gravity in the 5D braneworld.
  \item[b)] Smoothness of the curvature scalar $R_5.$
  \item[c)] Existence of a mass gap in the spectrum of gravitational excitations of the system.
\end{description}
From the expressions for $\Psi_{0}$ (\ref{Psi0}), $V_{QM}$
(\ref{VQM}) and $R_5$ (\ref{R}) we find that the curvature scalar
can be written in terms of the zero energy wave function and the
quantum mechanical potential as follows:
\begin{equation}
R_5=\frac{16}{3}\Psi_{0}^{-\frac{4}{3}}V_{QM}. \label{relation}
\end{equation}
Since the warp factor is real, (\ref{Psi0}) and (\ref{norma}) imply
that $\Psi_{0}$ must tend to zero asymptotically
\begin{equation}
\Psi_{0}|_{z\rightarrow\infty}\rightarrow 0 \label{Psi0atinfty}
\end{equation}
in order to ensure 4D gravity localization. If we indeed require the
existence of a mass gap, $V_{QM_\infty}$ must adopt a positive value
asymptotically.

Recall that we are considering thick brane configurations that are
regular at the position of the brane. Thus, if a) and c) are
fulfilled, (\ref{relation}) implies that $R_5$ will necessary
possess naked singularities as $|z|\rightarrow\infty.$ On the other
side, if we assume that conditions a) and b) are satisfied, then
$V_{QM}$ must vanish asymptotically at least with the same rapidity
of $\Psi_{0}^{-\frac{4}{3}},$ implying, in turn, that there is no
mass gap for such a solution of the model.

Thus, if one desires to construct a completely regular thick brane
configuration in the framework of the model (\ref{action}) with the
ansatz (\ref{conflinee}) in which 4D gravity is localized, then the
corresponding mass spectrum will have no mass gap between the zero
mode bound state, identified with a stable 4D graviton, and the
continuous spectrum of massive KK excitations.

Conversely, if we require the existence of a mass gap in such a
model restricted by (\ref{conflinee}), then the scalar curvature
will necessary develop naked singularities asymptotically. This
singularities can be made harmless if one imposes unitary boundary
conditions in the spirit of \cite{gremm},\cite{GMZ} and
\cite{cohenkaplan}, providing a viable model from the physical point
of view.

The fact that the manifold develops naked singularities at the
boundaries in this case does not matter if no conserved quantities
are allowed to leak out through the boundaries. The 4D Poincar\'e
isometries of metric (\ref{conflinee}) corresponds to 4D
conservation laws. Thus, by considering the translations generated
by the Killing vectors $\xi^\mu_m=\delta^\mu_m$, where $\mu$ is a 5D
index, one can construct currents by contracting with the stress
tensor \cite{gremm}
\begin{equation}
J^\mu = T^{\mu\nu}\xi^{m}_\nu. \label{current}
\end{equation}
These currents satisfy a covariant conservation law of 4D energy and
momentum
\begin{equation}
\frac{1}{\sqrt{g}}\partial_\mu\left(\sqrt{g} J^\mu\right) = 0.
\label{conservationlaw}
\end{equation}
Therefore we demand that the flux through the singular boundary of
spacetime (along the transverse direction) must vanish for all
currents in order to ensure that these quantities are conserved in
the presence of a singularity:
\begin{equation}
\lim_{z\to\infty} \sqrt{g} J^z = \lim_{z\to\infty} \sqrt{g} g^{zz}
\frac{1}{2} \partial_m h^T_{pq}\partial_z h^T_{pq} = 0.
\label{noflux}
\end{equation}

We shall see below that this unitary boundary conditions project out
all continuum modes from the spectrum of KK graviton excitations,
leading to a unitary spectrum consisting of discrete bound states in
the theory.

\subsection{Solution with a mass gap}

The system under consideration allows for the following solution
\cite{dewolfe,gremm}
\begin{eqnarray}
\label{sol} e^{2A}=\left[\cos\left(ay-y_0\right)\right]^b,\qquad
\phi=\frac{\sqrt{3b}}{2}\ln\left[\sec\left(ay-y_0\right)+\tan\left(ay-y_0\right)\right],\\
V(\phi)=\frac{3a^2b^2}{4}\left[1+\frac{1-2b}{b}
\cosh^2\left(\frac{2\phi}{\sqrt{3b}}\right)\right].\nonumber
\end{eqnarray}
where $a, b$ and $y_0$ are arbitrary constants.

Such a solution describes a thick brane located at $y_0$ when the
range of the fifth dimension is $-\pi/2\le a(y-y_0)\le\pi/2$
(see \cite{bh2}) and involves naked singularities at the boundaries of the
manifold \cite{gremm,bhrs}; $a$ represents the inverse of the
brane's width ($\Delta\sim 1/a$) when $b>0$.

If $b=2$ one can invert the coordinate transformation
(\ref{coordtransf}) yielding the following relation:
$$
\cos\left[a(y-y_0)\right]={\rm sech}(az).
$$
This transformation decompactifies the fifth dimension and sends the
naked singularities that were present at $-\pi/2a$ and $\pi/2a$ to
spatial infinity \cite{gremm,bhrs}. In terms of $z$, the warp factor
becomes
$$A(z)=\ln{{\rm sech}(az)},$$
and the analog quantum mechanical potential adopts the form
\cite{bhrs}
\begin{equation}
\label{potentialpes3e4} V_{QM}(z)=\frac{3a^2}{4}\{3-5{\rm
sech}^2(az)\}.
\end{equation}
Recall that the eigenvalue spectrum denoted by $m^2$ parameterizes
the spectrum of graviton masses that a 4D observer standing at $z=0$
sees. The analog quantum mechanical potential asymptotically
approaches a positive value $V_{QM}(\infty)=9a^2/4$, leading to a
mass gap between the zero mode bound state and the first KK massive
mode \cite{csakietal}. By setting $u=az$ the Schr\"odinger equation
can be transformed into the standard form possessing a modified
P\"oschl--Teller potential with $n=3/2$ (see
\cite{bhrs},\cite{PTptl})
\begin{equation}
\label{schrodingerPT} \left[-\partial_u^2-n(n+1){\rm
sech}^2u\right]\Psi(u)=
\left[\frac{m^2}{a^2}-\frac{9}{4}\right]\Psi(u)=E\Psi(u).
\end{equation}
Thus, the wave function possesses two bound states: the ground state
$\Psi_0$ with energy $E_0$ and an excited massive state $\Psi_1$
with energy $E_1$. In this case the Schr\"odinger equation
(\ref{schrodinger}) can be solved analytically for arbitrary $m$ and
the general solution is a linear combination of associated Legendre
functions of first and second kind of degree $3/2$ and order
$\mu=\sqrt{\frac{9}{4}-\frac{m^2}{a^2}}$: \be \Psi_m=k_1\
P_{3/2}^{\mu} \left[\tanh(az)\right]+k_2\
Q_{3/2}^{\mu}\left[\tanh(az)\right],\ee where $k_1$ and $k_2$ are
integration constants. The ground state corresponds to the massless
state $m=0$ ($\mu=3/2$) and possesses energy $E_0=-n^2=-9/4$,
whereas the excited state has mass $m=\sqrt{2}a$ ($\mu=1/2$) and
energy $E_1=-(n-1)^2=-1/4$ .

Since for the special values $\mu=3/2$ and $\mu=1/2$ the associated
Legendre function's series are finite, one obtains the following
eigenfunction for the massless mode \be \Psi_0=C_0\ {\rm
sech}^{3/2}(az), \label{psi0}\ee whereas the excited massive mode
adopts the form \be \Psi_1=C_1\ {\rm sinh}(az){\rm sech}^{3/2}(az),
\label{psi1}\ee where $C_0$ and $C_1$ are arbitrary normalization
constants. Since (\ref{psi0}) represents the lowest energy
eigenfunction of the Schr\"odinger equation (\ref{schrodinger}) and
it has no zeros, it can be interpreted as a massless 4D graviton
with no tachyonic instabilities from modes with $m^2<0$. On the
other side, the eigenfunction (\ref{psi1}) represents a normalizable
massive graviton localized on the center of the thick brane. The
mass gap existing between these two bound states eliminates the
potentially dangerous arbitrarily light KK excitations, sending them
to energies of the scale of $a$.

In the spectrum there is also a continuous part of massive modes
with $2m>3a$ (the order becomes purely imaginary $\mu=i\rho$, where
$\rho=\sqrt{(m/a)^2-9/4})$ that describe plane waves as they
approach spatial infinity \cite{bhrs} and give rise to small
corrections to Newton's law in 4D flat spacetime in the thin brane
limit \cite{csakietal,bhnq}.

However, by imposing the unitary boundary conditions mentioned
above, i.e., by demanding the vanishing of the flux through the
singular boundary of our spacetime:
\begin{equation}
\lim_{z\to\infty}
e^{3A/2}\Psi(z)\partial_z\left[e^{-3A/2}\Psi(z)\right]\sim
\Psi(z)\left[(3B-2A\rho)\cos(a\rho z)+(3A+2B\rho)\sin(a\rho
z)\right] = 0, \label{noflux}
\end{equation}
where we have made use of the asymptotic form of $\Psi(z):$
\begin{equation}
\Psi(z)\sim A\sin(a\rho z)+B\cos(a\rho z),
\end{equation}
we get a result implying that $A$ and $B$ must vanish, just
eliminating all the continuum modes from the spectrum of KK
excitations. Thus, the unitary spectrum consists of two discrete
modes because they do not generate any flux into the singularity.

\section{Conclusions and discussion}

We have a 5D thick braneworld arising in a scalar tensor theory and
have investigated the relationship existing among the localization
of 4D gravity, the smooth or singular character of the curvature
scalar of the classical background, and the existence of a mass gap
which separates the massless graviton bound state from the massive
KK excitations. We have shown that when considering 4D gravity
localization in the 5D thick braneworld, the smooth character of the
scalar curvature of the manifold excludes the possibility of the
existence of a mass gap in the graviton spectrum of fluctuations;
conversely, by demanding the existence of a mass gap between the
massless and massive modes of KK excitations, the curvature scalar
necessarily develops naked singularities at the boundaries of the
manifold. We further considered an explicit solution which gives
rise to a mass gap in the spectrum of graviton fluctuations. This
solution contains a massless graviton, one massive excited state
with $m= \sqrt2 a$, and a continuum of modes with a mass gap of size
$m= 3a/2$. At very low energies, none of these massive modes can be
excited, and an observer at $z=0$ sees pure 4D gravity. At higher
energies the massive state can be excited, giving some corrections
to Newton's law and, finally, at energies larger than the gap the
whole continuum of modes can be excited. Violations of unitarity
occur only when modes that can travel out to the singularities can
be excited, i.e.~only at energies above the mass of the lightest
continuum mode.

For this reason we imposed unitary boundary conditions on the
described solution, a fact that eliminates all the continuous modes
from the spectrum of KK excitations of the classical background and
renders the naked singularities harmless (in the same spirit as in
\cite{gremm},\cite{GMZ},\cite{cohenkaplan}).

\section{Acknowledgements}

This research was supported by grants CIC--4.16 and CONACYT
60060--J. DMM and RRML acknowledge a PhD grant from CONACYT and
UMSNH, respectively. AHA and UN thanks SNI for support.

\end{document}